\documentclass[aps,prl,floatfix,twocolumn,showpacs,showkeys,superscriptaddress]{revtex4-1}

\usepackage{graphicx}% Include figure files
\usepackage{dcolumn}% Align table columns on decimal point
\usepackage{bm}% bold math
\usepackage{SIunits}% bold math
\usepackage{color} % corlored text
\usepackage{ulem} % n\"{o}tig f\"{u}r's durchstreichen

\begin{document}

\preprint{}

\title{High cooperativity in coupled microwave resonator ferrimagnetic insulator hybrids}

\author{Hans Huebl}
\email[corresponding author
]{huebl@wmi.badw.de}\affiliation{Walther-Mei{\ss}ner-Institut, Bayerische Akademie der Wissenschaften, Garching, Germany}

\author{Christoph W. Zollitsch}
\affiliation{Walther-Mei{\ss}ner-Institut, Bayerische Akademie der Wissenschaften, Garching, Germany}

\author{Johannes Lotze}
\affiliation{Walther-Mei{\ss}ner-Institut, Bayerische Akademie der Wissenschaften, Garching, Germany}

\author{Fredrik Hocke}
\affiliation{Walther-Mei{\ss}ner-Institut, Bayerische Akademie der Wissenschaften, Garching, Germany}

\author{Moritz Greifenstein}
\affiliation{Walther-Mei{\ss}ner-Institut, Bayerische Akademie der Wissenschaften, Garching, Germany}

\author{Achim Marx}
\affiliation{Walther-Mei{\ss}ner-Institut, Bayerische Akademie der Wissenschaften, Garching, Germany}

\author{Rudolf Gross}
\affiliation{Walther-Mei{\ss}ner-Institut, Bayerische Akademie der Wissenschaften, Garching, Germany}
\affiliation{Physik-Department, Technische Universit\"{a}t M\"{u}nchen, Garching, Germany}

\author{Sebastian T. B. Goennenwein}
\affiliation{Walther-Mei{\ss}ner-Institut, Bayerische Akademie der Wissenschaften, Garching, Germany}

\date{\today}

\begin{abstract}
We report the observation of strong coupling between the exchange-coupled spins in gallium-doped yttrium iron garnet and a superconducting coplanar microwave resonator made from Nb. The measured coupling rate of $450\,\mega\hertz$ is proportional to the square-root of the number of exchange-coupled spins and well exceeds the loss rate of $50\,\mega\hertz$ of the spin system. This demonstrates that exchange coupled systems are suitable for cavity quantum electrodynamics experiments, while allowing high integration densities due to their spin densities of the order of one Bohr magneton per atom. Our results furthermore show, that experiments with multiple exchange-coupled spin systems interacting via a single resonator are within reach.
\end{abstract}

\pacs{85.25.-j, 85.70.Ge, 42.50.Pq, 76.30.+g}

\keywords{ferromagnetic resonance, ferrimagnetic resonance, strong coupling, cavity quantum electrodynamics, YIG, yttrium iron garnet, low temperatures, microwave spectroscopy}
\maketitle

The study of the interaction of matter and light on the quantum level is at the core of solid state quantum information systems. Strong \cite{wallraff04, schoelkopf08} and ultra-strong coupling \cite{Niemczyk:2010gv} has been achieved, allowing for the coherent transfer of quantum information. For the practical implementations of quantum information systems, the use of hybrid systems has been suggested. In such hybrids,  natural microscopic systems (atoms, molecules, electron spins, and nuclear spins) are coupled with artificial meso-scale structures such as superconducting quantum circuits by means of microwave photons \cite{Andre:2006, Verdu:2009, Wallquist:2009}. Whereas the former have long coherence times due to sufficient decoupling from environmental noise, the latter allow for fast qubit gates due to strong coupling to electromagnetic fields \cite{dicarlo09}.  Ensembles of electron spins as quantum memories \cite{Wesenberg:2009es, Imamoglu:2009eg} seem promising and their coupling to superconducting resonators \cite{Schuster:2010bu, Kubo:2010iq, Kubo:2011dx, Amsuss:2011ci, Chiorescu:2010hw, Bushev:2011be, Abe:2011gv} and flux qubits \cite{Zhu:2011} has been studied recently. Although the coupling strength $g$ of an individual spin to the electromagnetic mode of a superconducting microwave resonator is small (typically $10\,\hertz$), the coupling of an ensemble of $N$ spins is enhanced by a factor of $\sqrt{N}$ \cite{Raizen:1989, wallsmilburn94}. In this way, strong coupling $g_{\rm{eff}}= g\sqrt{N} \gg \kappa, \gamma$ can be realized, where $\kappa$ and $\gamma$ are the loss rates of the resonator and spin system, respectively. With loss rates in the order of MHz, typically 10$^{12}$ spins are needed to reach the strong coupling regime. Until today, mostly paramagnetic systems consisting of ensembles of noninteracting spins have been studied. The coherent coupling of microwave resonators to ferromagnetic systems with strongly exchange coupled spins remains to be explored. Soykal and Flatt\'{e} \cite{Soykal:2010er, Soykal:2010hz} theoretically discussed the strong coupling of photonic and magnetic modes in exchange locked ferromagnetic systems. Two particular advantages of ferromagnetic systems are (i) their higher spin density, such that for the same number $N$ of spins, their volume can be reduced considerably compared to dilute paramagnetic systems, and (ii) the fact that below the magnetic ordering (Curie) temperature  the system essentially is fully polarized, in contrast to the thermal polarization in uncoupled spin ensembles. This should allow to couple multiple spin ensembles to the same microwave resonator, e.g. for realizing an adjustable coupling between the magnetic subsystems or for the exchange of individual quanta between them.

In this letter, we investigate the coupling between the electromagnetic modes of a superconducting coplanar waveguide microwave resonator and the magnetic modes of the exchange-locked ferrimagnet yttrium iron garnet (Y$_3$Fe$_5$O$_{12}$ or YIG) doped with gallium (YIG:Ga). We measure a coupling rate of $g_{\rm{eff}}/2 \pi = 450\,\mega \hertz$ exceeding both the spin relaxation rate $\gamma/2\pi= 50\,\mega\hertz$ and the resonator decay rate $\kappa/2\pi= 3\,\mega\hertz$. That is, we observe strong coupling. The measured effective coupling strength follows $g_{\rm{eff}}= g\sqrt{N}$, where the number of spins interacting with the resonator is estimated from the sample geometry. Furthermore, the measured relaxation rate of the spin system is fully consistent with the natural linewidth of YIG:Ga obtained from ferromagnetic resonance (FMR) measurements \cite{Rachford:2000en}. Pure YIG is one of the prime candidates for studying strong coupling between exchange locked spins and the electromagnetic modes of a microwave resonator, in particular because of its very small FMR linewidth of $\approx 10\,\micro\tesla$ at $4\,\kelvin$ and $\omega /2\pi = 9.3\,\giga\hertz$ \cite{SPENCER:1959wv}. This narrow linewidth corresponds to a $T_2$ time in the order of microseconds \cite{Kaplan:1965hf}. Since high quality YIG thin films can be prepared on various substrates (gadolinium gallium garnet \cite{Manuilov:2009kl, Manuilov:2010ix}, Si and GaAs \cite{Levy:1997ki}) and doped with rare earth elements in order to adjust the FMR linewidth in a controlled way, YIG seems an ideal material for ferromagnet based quantum hybrids.

\begin{figure}
\includegraphics[width=0.8\columnwidth]{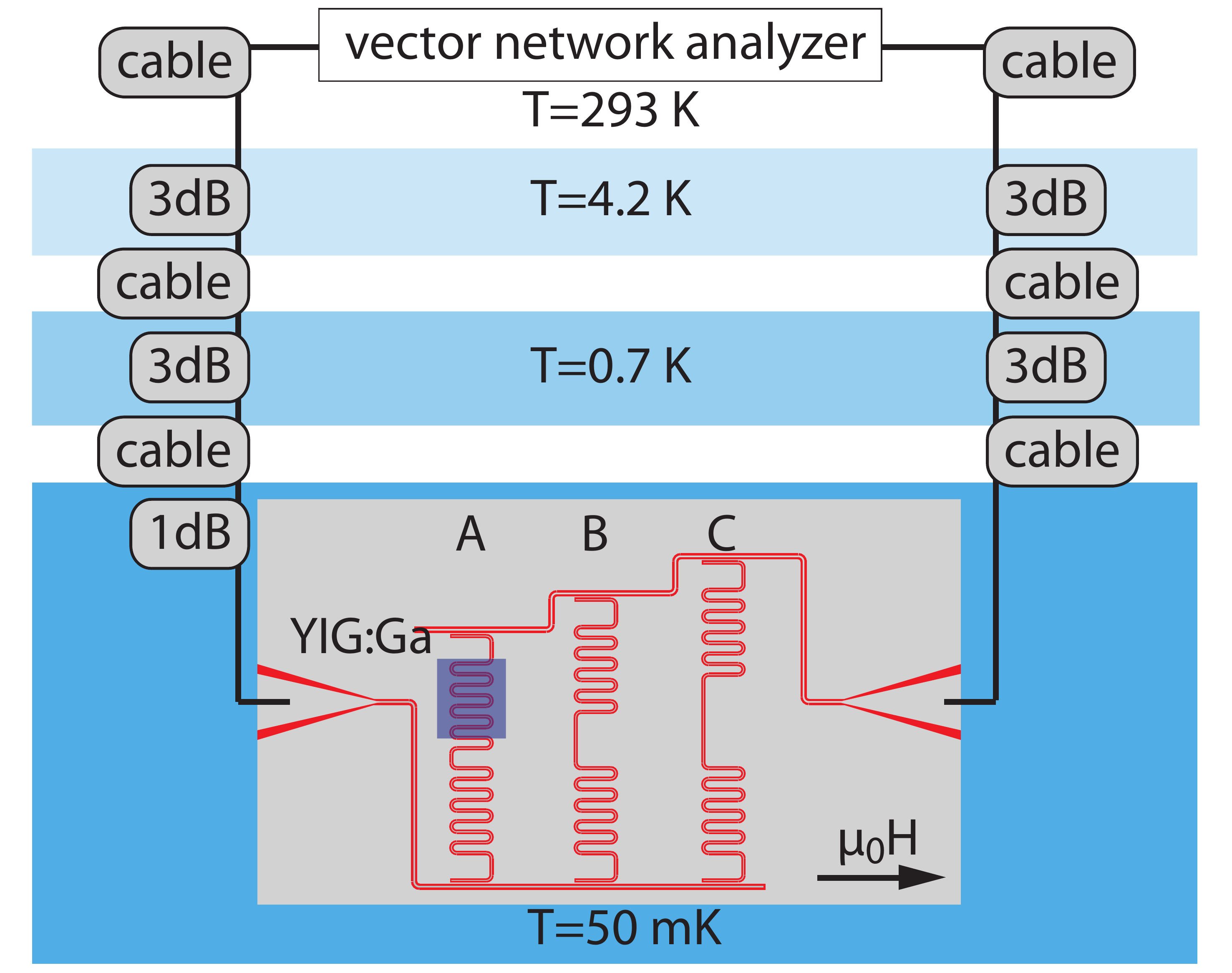}
\caption{Schematic of the experimental setup. The (purple) gallium doped YIG sample is cemented on top of one of the niobium microwave resonators which are arranged to allow for multiplexing. Experiments are performed at millikelvin temperatures in transmission by vector network analysis in a superconducting solenoid magnet.
 \label{fig:sample}}
\end{figure}

As pointed out by Soykal and Flatt\'{e} \cite{Soykal:2010er, Soykal:2010hz}, in a macrospin approximation the Hamiltonian for the ferromagnet-resonator system can be expressed as
\begin{equation}
\mathcal{H} =\hbar\omega_r a^\dag a  + g_s \mu_{\rm B} B_z^{\rm{eff}} S_z + \hbar g (a S_+ + a^\dag S_-) .
\label{Hamiltonian}
\end{equation}
Here, $a^\dag$ and $a$ are the photon creation/annihilation operators, $\omega_r$ the resonator frequency, $g_s$ the electron $g$-factor, $\mu_{\rm B}$ the Bohr magneton, and $B_z^{\rm{eff}}$ the magnetic field \footnote{The magnetic induction $B_z^{\rm{eff}}=B_z^{\rm{ext}}+B_{\rm{a}}$ has two contributions: i) the externally applied magnetic field $B_z^{\rm{ext}}$ and ii) internal fields accounting for the magnetic anisotropy $B_{\rm{a}}$. (cf. \cite{SI})}. The macrospin operator $\mathbf{S}= ( S_+ \hat{\mathbf{e}}_- -  S_- \hat{\mathbf{e}}_+)/\sqrt{2} + S_z\hat{\mathbf{z}}$ with  $\hat{\mathbf{e}}_\pm = \mp (\hat{\mathbf{x}} \pm \imath\hat{\mathbf{y}})/\sqrt{2}$ is expressed in terms of the spin lowering and raising operators
\begin{equation}
S_\pm \left| \frac{N}{2}, m \right\rangle = \sqrt{\left(\frac{N}{2} \mp m\right)\left(\frac{N}{2} \pm m+1 \right)} \; \left|\frac{N}{2}, m\pm 1 \right\rangle .
\label{spin_lowering_raising}
\end{equation}
Here, $| \ell, m \rangle $ are the eigenstates of the macrospin and we have assumed that the macrospin state is fixed at its maximal value $\ell = N/2$. In the Dicke model \cite{Dicke:1954} of $N$ independent paramagnetic spins this would correspond to a fully excited spin system with no photons in the cavity. In contrast to the Dicke model, for our macrospin model states with $\ell < N/2$ are not accessible. Due to strong exchange coupling, states with $\ell < N/2$ are separated in energy and require the excitation of magnons. We note that the coupling between the photonic and magnetic system is a magnetic dipole transition and that the Hamiltonian (\ref{Hamiltonian}) conserves the total excitation number $Z = n + m$, where $n$ is the photon number in the cavity and $|m| \le \ell = N/2$ the magnetic quantum number. Assuming that $\mathbf{S}$ is antiparallel to $B_z$ and $n=0$ initially, we have $Z = N/2$ and, hence, can index the basis states $|n, m\rangle$ of the resonator-spin systems either by the photon number $n$ ($|n, \frac{N}{2}-n\rangle$) or the magnetic quantum number $m$ ($|\frac{N}{2}-m, m\rangle$). Evidently, these basis states are similar to those of the Dicke model \cite{Dicke:1954} for a paramagnetic ensemble of $N$ noninteracting spins coupled to a resonator, with $\ell = N/2$ taking the role of the cooperation number.

Due to the analogy with the Dicke model, we expect that the coupling strength of the ferromagnet-resonator system is given by the effective coupling strength $g_{\rm{eff}}= g\sqrt{N}$ of a paramagnet-resonator system, where $g=\frac{g_s \mu_B}{2\hbar}\,B_{1,0}$ is the coupling rate of an individual spin with the magnetic quantum number $m=1/2$ to the resonator \cite{Dicke:1954}. It is determined by the magnetic component of the rf vacuum field $B_{1,0} = \sqrt{\mu_0 \hbar \omega_{\rm{r}}/2 V_{\rm{m}}}$ which depends on the resonance frequency of the microwave resonator $\omega_{\rm{r}}$ and its mode volume $V_{\rm{m}}$ \cite{niemczyk09}. Hence, for a given resonance frequency, $g_{\rm{eff}}=\frac{g_s\mu_B}{2\hbar}\sqrt{\mu_0 \rho \hbar \omega_{\rm{r}}V/2V_{\rm{m}}}$ depends only on the spin density $\rho =N/V$ and the filling factor $V/V_{\rm{m}}$, where $V$ denotes the volume of the resonator field mode filled with the spin system. Non-interacting spin ensembles like paramagnetic centers in semiconductors or insulators typically have a spin density $\rho$ in the order of $10^{15}\leq \rho \leq 10^{18}\,\centi\meter^{-3}$ \cite{Amsuss:2011ci, Schuster:2010bu, Kubo:2010iq}. For $\omega_{\rm r}/(2\pi)\approx 5\,\giga\hertz$, this results in coupling rates in the order of $10\,\mega\hertz$ assuming $V/V_{\rm{m}}\simeq 1$. In contrast, exchange coupled systems naturally have a spin density in the order of one per atom (e.g. Fe, Ni, Co) or in the case of YIG $40$ per unit cell (unit cell volume - $1.8956\,\nano\meter^3$), corresponding to a spin density of  $2\times10^{22}\,\rm{cm^{-3}}$ \cite{gilleo58}. Due to the increase of at least four orders of magnitude in the spin density we expect  a two orders larger coupling strength in exchange coupled systems as compared to non-interacting spins.  Therefore, the exchange coupled system sample volume  can be reduced by a factor of $10^4$ while keeping the coupling rate constant, enabling a higher integration density.

In our experiments, we study the coupling between the exchange locked system YIG:Ga and a superconducting Nb resonator. The resonator structure is patterned into a $100\,\nano\meter$ thick Nb film deposited onto an intrinsic silicon substrate using optical lithography and reactive ion etching \cite{niemczyk09}. Figure~\ref{fig:sample} shows the layout of the microwave circuitry consisting of an input line, three  resonators with resonance frequencies of $f_A=5.90\,\giga\hertz$, $f_B=5.53\,\giga\hertz$, and $f_C=5.30\,\giga\hertz$, and an output line. This configuration allows to compare loaded and unloaded microwave resonators on the same chip. A $2 \times 0.5 \times 0.7 \milli\meter^3$ sized (length $\times$ width $\times$ thickness) commercial YIG:Ga crystal is cemented onto resonator $A$ with the highest microwave frequency $f_A=5.90\,\giga\hertz$. The number of spins interacting with the resonator is roughly estimated from the overlap of the YIG:Ga crystal with the meandering coplanar waveguide with a center conductor width of $6~\micro\meter$ and a gap of $12~\micro\meter$.  With the overlap length of $2.5~\milli\meter$ and assuming that the vertical extension of the microwave field into the YIG crystal is about $30~\micro\meter$, the total number of spins coupled to the resonator is estimated to $N\approx 4.5\times 10^{16}$. To preserve the superconducting state of the microwave resonators, the surface of the chip is carefully aligned in parallel to the applied magnetic field $B_z$ generated by a  superconducting solenoid. The microwave transmission experiments are performed at the base temperature of a dilution refrigerator of $50\,\milli\kelvin$ using a commercial vector network analyzer. To thermally anchor the center conductor of the  microwave input and output lines, attenuators are used at the $4\,\kelvin$, the still and the mixing chamber stages (cf.\,Fig.\,\ref{fig:sample}). Considering only the attenuators, we estimate a microwave field temperature of about $70\,\kelvin$ (or 290 thermally excited photons on average) in the resonator. \footnote{This accounts for the fixed attenuators and does not include the lossy lines. For further details refer to \cite{SI}.}

\begin{figure}
\includegraphics[width=0.9\columnwidth]{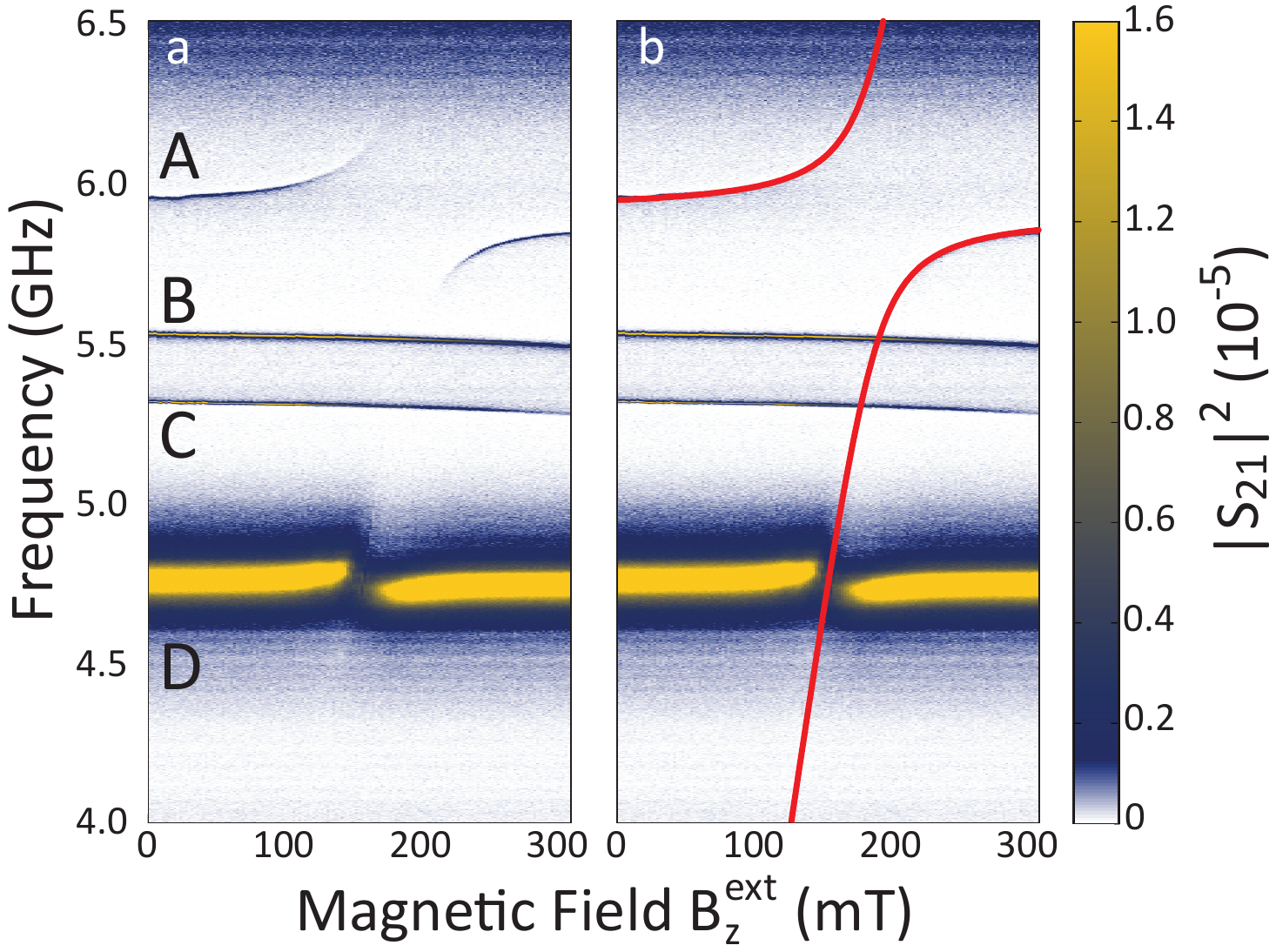}
\caption{Transmission spectrum of the setup including the YIG-microwave resonator hybrid  as a function of the applied magnetic field $B_z^{\textrm{ext}}$, taken at  $T=50\,\milli\kelvin$ [33]. The coplanar waveguide resonators ($A$-$C$) show a slightly decreasing resonance frequency with increasing in-plane magnetic field. Additionally, resonator $A$ shows a pronounced avoided crossing at $170\,\milli\tesla$, where the resonator frequency $f_A$ matches the FMR frequency $\omega_{\rm FMR}/(2\pi)$. Resonance $D$ is a parasitic mode present in the sample box. Panel a) shows the raw (uncalibrated) transmission data as measured. Panel b) shows the same data again superimposed with a fit according to eq.(\ref{two_level_system}) plotted as red line.
 \label{fig:spectra}}
\end{figure}

\begin{figure}
\includegraphics[width=0.9\columnwidth]{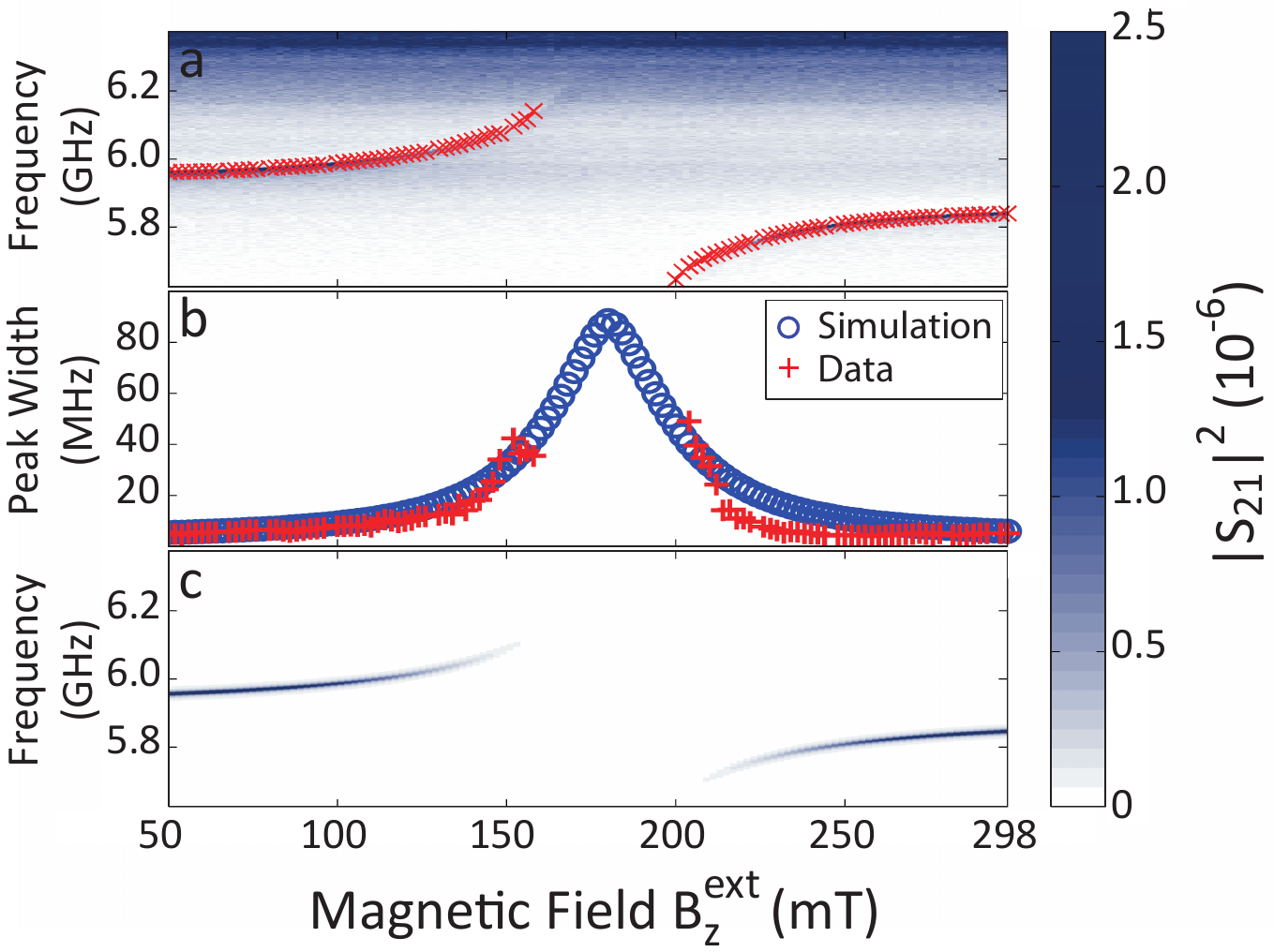}
\caption{Analysis of the resonance frequency and linewidth as a function of the external magnetic field $B_z^{\textrm{ext}}$. Panel (a) shows the resonator transmission $|S_{21}|^2$ (same data as Fig.\,\ref{fig:spectra}) as a function of frequency and applied magnetic field [33]. The red crosses mark the resonance frequency determined by fitting a Lorentzian lineshape to the data at constant $B_z^{\textrm{ext}}$. The red crosses in panel (b) show the extracted FWHM linewidths corresponding to $2\gamma/(2\pi)$ and $2\kappa/(2\pi)$. In addition, the blue circles show the linewidths obtained from the numerical simulation of the transmission spectra plotted in panel (c). The simulation is based on the input-output formalism resulting from eq.(\ref{inputoutput}) \cite{Schuster:2010bu, Abe:2011gv, Clerk:2010dh, wallsmilburn94}.
 \label{fig:analysis}}
\end{figure}

Figure~\ref{fig:spectra} shows the microwave transmission $\left|S_{21}\right|^2$ raw data as a function of frequency and applied magnetic field \footnote{Since we have not calibrated our setup at millikelvin temperatures, we show raw, uncalibrated, as-measured $S_{21}$ transmission data in Figs.\,\ref{fig:spectra} and\,\ref{fig:analysis}. In our opinion, this is the most honest way of presenting the data in lack of a proper full calibration. \label{footnote:uncalibrated}}. In the spectrum at $B_z=0$, four transmission peaks are visible corresponding to the resonance frequencies of the coplanar microwave resonators $A$, $B$, and $C$. The broad feature labeled $D$ stems from a parasitic mode of the metallic microwave box in which the sample is mounted. As expected, the resonators $B$ and $C$ show only a weak magnetic field dependence, because they are not interacting with the YIG:Ga crystal due to the absence of physical overlap (cf.~Figs.~\ref{fig:sample} and \ref{fig:spectra}). On the contrary, resonator $A$ and the box mode $D$ couple to the YIG:Ga. While mode $D$ allows us to probe the ferromagnetic resonance independently of the strongly coupled mode ($A$) and to determine the FMR dispersion relation, resonator $A$ shows a distinct anticrossing at $B_z^{\rm{ext}}(\Delta=0)=B_\textrm{FMR} = 170\,\milli\tesla$ where the FMR dispersion relation $\hbar\omega_\textrm{FMR} = g_s \mu_{\rm B} B_z^{\rm{eff}}$ \cite{SI} is degenerate with the resonator $\hbar\omega_{\rm r}$.

To derive the effective coupling rate $g_{\rm{eff}}$ from the measured data, we simplify the discussion of (\ref{Hamiltonian}), by modeling the system as two coupled harmonic oscillators. The dispersion of the resonance frequency is then given by \cite{HarocheRaimond:2006}
\begin{equation}
\omega=\omega_{\rm{r}} + \frac{\Delta}{2} \pm \frac{1}{2}\,\sqrt{\Delta^2 + 4 g_{\rm{eff}}^2} \, .
 \label{two_level_system}
\end{equation}
Here, $\Delta= \omega_{\rm{FMR}} - \omega_{\rm{r}} = g_s\mu_{\rm{B}}(B_{z}^{\textrm{ext}} - B_{\rm{FMR}}) / \hbar$ is the field dependent detuning between the resonator frequency $\omega_{\rm{r}}/(2\pi) = f_A$ and the field dependent FMR frequency $\omega_{\rm{FMR}}/(2\pi)$. The experimental data agree very well with this model prediction. Fitting the data yields $g_{\rm{eff}}=450\pm20\,\mega\hertz$ and $B_{\rm{FMR}}=170\pm5\,\milli\tesla$, and the $g$-factor of the ferrimagnetic resonance $g_s=2.17\pm0.05$ (red line in Fig.\ref{fig:spectra}). Here, $B_{\rm{FMR}}$ is reduced with respect to the bare electron spin resonance field of 194 mT due to the presence of an anisotropy field $B_a=24\,\milli\tesla$ [28],\cite{SI}. Additionally, the experimentally observed $g_{s}$ is not exactly identical to the literature value for pure YIG at 2\,K of $g_{s,\rm{lit}} = 1.99$ \cite{Belov:1960va} or YIG:Ga at 5\,K of $g_{s,\rm{lit}} = 2.1$\cite{Rachford:2000en}. The observed difference might be due to the higher Ga concentration compared to Ref.\,\cite{Rachford:2000en} or the lower temperature. Note, that the $g$-factor and the magnetic anisotropy of YIG and doped YIG is not well established and requires further investigations. For our resonator we estimate $g/(2\pi) \simeq 5\,\hertz$ \cite{niemczyk09} and $N \simeq 4\times 10^{16}$. With these numbers we expect $g_{\rm{eff}}/(2\pi)=(g/2\pi)\, \sqrt{N} \simeq 1\,\giga\hertz$ corroborating our experimental result within a factor of two despite of the rough estimate for $N$.

To determine the relaxation rate $\gamma$ of the spin system we analyze the evolution of the linewidth of the resonator mode $A$ as a function of the magnetic field $B_z$ by fitting a Lorentzian lineshape in the frequency domain for every measured magnetic field magnitude \cite{Abe:2011gv}. Figure~\ref{fig:analysis}(a) shows the resonance frequency obtained from such a fit as red crosses superimposed on the color-coded dataset. In Fig.~\ref{fig:analysis}(b) the corresponding (FWHM) linewidth (red crosses) is shown. At low magnetic fields, the resonator mode $A$ is essentially decoupled from the spin system,  such that the measured linewidth is given by $\kappa$. Closer to the ferromagnetic resonance, the linewidth of the system is given by the combined relaxation rate of the spin system and the microwave resonator leading to an increase in the observed linewidth.

To quantify the coupling and loss rates in our system, we use a standard input-output formalism \cite{Schuster:2010bu, Abe:2011gv, Clerk:2010dh, wallsmilburn94}. Within this framework the transmission amplitude of microwaves from the input to the output port of the microwave resonator is given by
\begin{equation}
S_{21} = \frac{\kappa_c}{\imath(\omega - \omega_{\rm{r}}) -(\kappa_{\rm{c}}+\kappa_{\rm{i}})+\frac{|g_{\rm{eff}}|^2}{\imath(\omega-\omega_{\rm{FMR}})-\gamma/2}} \; .
 \label{inputoutput}
\end{equation}
Here, $\omega/2\pi$ is the frequency of the microwave probe tone, $\kappa_{\rm{c}}$ is the external coupling rate between the microwave resonator and the feed line, and $\kappa_{\rm{i}}$ summarizes the intrinsic loss rate of the microwave resonator. In our case we have $\kappa_{\rm{i}} \gg \kappa_{\rm{c}}$, resulting in a total microwave resonator relaxation rate $\kappa \simeq \kappa_{\rm{i}}$ (cf. \cite{SI}). Fig.\,\ref{fig:analysis}(c) shows the calculated transmission using $g_{\rm{eff}}/2\pi=450\,\mega\hertz$, $\gamma/2\pi=50\,\mega\hertz$, and  $\kappa/2\pi=3\,\mega\hertz$. Evidently, all features of the experimental data of Fig.\,\ref{fig:analysis}(a) are nicely reproduced.  Moreover, the two transmission peaks expected at $B_{\rm{FMR}}=170\,\milli\tesla$ cannot be resolved due to the limited signal to noise ratio in the experimental data. Additonally, we can analyze the simulation data shown in Fig.\,\ref{fig:analysis}(c) in the same way as the experimental data in Fig.\,\ref{fig:analysis}(a).  The result is shown by the blue circles in Fig.\,\ref{fig:analysis}(b), where in contrast to the experimental data, the modeled data is noise-free allowing to predict the linewidth for all magnetic field values. The good agreement between experimental and simulation data again demonstrates that the parameters chosen in the simulation well reproduce the experimental situation. In summary, our analysis shows that $g_{\rm{eff}} \gg \kappa, \gamma$ with a cooperativity $C=g_{\rm{eff}}^2/\kappa\gamma \simeq 1350$. That is, the strong coupling regime has been reached for the ferrimagnet-resonator system \footnote{Our claims are based on the analysis of the lineshape and resonance frequency dependence as function of the applied magnetic field and do not rely on the absolute transmission intensity. We therefore have not calibrated the setup with respect to the amplitude information.}.

Next, we compare the experimentally determined relaxation rate $\gamma$ with the temperature dependence of the FMR linewidth measured at $9.43\,\giga\hertz$. Typically, YIG:Ga exhibits significantly larger damping (larger linewidth) than pure YIG. Rachford {\it et al.} \cite{Rachford:2000en} report linewidths for YIG:Ga that decrease from $1\,\milli\tesla$  at $4.2\,\kelvin$ to $0.1\,\milli\tesla$ at room temperature, corresponding to $28\,\mega\hertz$ and $2.8\,\mega\hertz$, respectively. Our sample has a higher Ga-doping concentration \cite{gayig-datasheet-2012}, and thus larger linewidth, which coroborates the relaxation rate $\gamma/2\pi=50\,\mega\hertz$ we measured at millikelvin temperatures.  However, note that little is known about the damping in YIG for $T\leq 2\,\kelvin$. According to Sparks and Kittel \cite{Sparks:1960} the limiting relaxation mechanism is spin-lattice coupling, expected to be well below $1\,\mega\hertz$ for YIG.  This calls for further experiments in this temperature regime to verify the proposed relaxation mechanism and to eludicate the maximum spin coherence time achievable in YIG for $T\leq2\,\kelvin$. Note also, that relaxation mechanisms in ferromagnets are fundamentally different from relaxation mechanisms in diluted paramagnetic systems. In the latter, spin-spin interactions can cause dephasing and decoherence \cite{Tyryshkin:2011fi}. In the former, it is possible to simultaneously have high spin density and low damping.

Finally, we find that $g_{\rm{eff}}$ is independent of the microwave power from $10\,\femto\watt$ to $10\,\nano\watt$.  This is expected, because here the number of excitations (photons in the microwave resonator) is much smaller than the number of spins $N\approx10^{16}$ \cite{Chiorescu:2010hw}. Only if the number of excitations (photons) becomes comparable to or exceeds $N$, a quenching of the observed anticrossing is expected.

In conclusion, we experimentally observed strong coupling between a superconducting microwave resonator and a gallium doped yttrium iron garnet ferrimagnet. The effective coupling rate of $450\,\mega\hertz$ reaches $8\%$ of the resonator frequency $\omega_{\rm r}/(2\pi)$ and by far exceeds the relaxation rate $\gamma/(2\pi)=50\,\mega\hertz$ of the spin system at about $50\,\milli\kelvin$, which is rather large owing to the gallium doping. Considering the much smaller linewidth in pure YIG, even higher cooperativities can be anticipated. Our results establish that exchange coupled spin systems indeed can be used for cavity quantum electrodynamics. Furthermore, the large coupling rates achievable in exchange coupled systems allow to place more than  one magnetic system in the microwave resonator. This should allow to study e.g. the exchange of magnetic excitations via a cavity bus similar to the approaches pursued in the field of cavity QED \cite{dicarlo09}.

We  acknowledge technical support by M. Opel. This work is supported by the German Research Foundation through SFB~631 and the German Excellence Initiative via the ``Nanosystems Initiative Munich'' (NIM).

\clearpage

\begin{thebibliography}{41}%
\makeatletter
\providecommand \@ifxundefined [1]{%
 \@ifx{#1\undefined}
}%
\providecommand \@ifnum [1]{%
 \ifnum #1\expandafter \@firstoftwo
 \else \expandafter \@secondoftwo
 \fi
}%
\providecommand \@ifx [1]{%
 \ifx #1\expandafter \@firstoftwo
 \else \expandafter \@secondoftwo
 \fi
}%
\providecommand \natexlab [1]{#1}%
\providecommand \enquote  [1]{``#1''}%
\providecommand \bibnamefont  [1]{#1}%
\providecommand \bibfnamefont [1]{#1}%
\providecommand \citenamefont [1]{#1}%
\providecommand \href@noop [0]{\@secondoftwo}%
\providecommand \href [0]{\begingroup \@sanitize@url \@href}%
\providecommand \@href[1]{\@@startlink{#1}\@@href}%
\providecommand \@@href[1]{\endgroup#1\@@endlink}%
\providecommand \@sanitize@url [0]{\catcode `\\12\catcode `\$12\catcode
  `\&12\catcode `\#12\catcode `\^12\catcode `\_12\catcode `\%12\relax}%
\providecommand \@@startlink[1]{}%
\providecommand \@@endlink[0]{}%
\providecommand \url  [0]{\begingroup\@sanitize@url \@url }%
\providecommand \@url [1]{\endgroup\@href {#1}{\urlprefix }}%
\providecommand \urlprefix  [0]{URL }%
\providecommand \Eprint [0]{\href }%
\providecommand \doibase [0]{http://dx.doi.org/}%
\providecommand \selectlanguage [0]{\@gobble}%
\providecommand \bibinfo  [0]{\@secondoftwo}%
\providecommand \bibfield  [0]{\@secondoftwo}%
\providecommand \translation [1]{[#1]}%
\providecommand \BibitemOpen [0]{}%
\providecommand \bibitemStop [0]{}%
\providecommand \bibitemNoStop [0]{.\EOS\space}%
\providecommand \EOS [0]{\spacefactor3000\relax}%
\providecommand \BibitemShut  [1]{\csname bibitem#1\endcsname}%
\let\auto@bib@innerbib\@empty
%</preamble>
\bibitem [{\citenamefont {Wallraff}\ \emph {et~al.}(2004)\citenamefont
  {Wallraff}, \citenamefont {Schuster}, \citenamefont {Blais}, \citenamefont
  {Frunzio}, \citenamefont {Huang}, \citenamefont {Majer}, \citenamefont
  {Kumar}, \citenamefont {Girvin},\ and\ \citenamefont
  {Schoelkopf}}]{wallraff04}%
  \BibitemOpen
  \bibfield  {author} {\bibinfo {author} {\bibfnamefont {A.}~\bibnamefont
  {Wallraff}}, \bibinfo {author} {\bibfnamefont {D.~I.}\ \bibnamefont
  {Schuster}}, \bibinfo {author} {\bibfnamefont {A.}~\bibnamefont {Blais}},
  \bibinfo {author} {\bibfnamefont {L.}~\bibnamefont {Frunzio}}, \bibinfo
  {author} {\bibfnamefont {R.-S.}\ \bibnamefont {Huang}}, \bibinfo {author}
  {\bibfnamefont {J.}~\bibnamefont {Majer}}, \bibinfo {author} {\bibfnamefont
  {S.}~\bibnamefont {Kumar}}, \bibinfo {author} {\bibfnamefont {S.~M.}\
  \bibnamefont {Girvin}}, \ and\ \bibinfo {author} {\bibfnamefont {R.~J.}\
  \bibnamefont {Schoelkopf}},\ }\href@noop {} {\bibfield  {journal} {\bibinfo
  {journal} {Nature}\ }\textbf {\bibinfo {volume} {431}},\ \bibinfo {pages}
  {162} (\bibinfo {year} {2004})}\BibitemShut {NoStop}%
\bibitem [{\citenamefont {Schoelkopf}\ and\ \citenamefont
  {Girvin}(2008)}]{schoelkopf08}%
  \BibitemOpen
  \bibfield  {author} {\bibinfo {author} {\bibfnamefont {R.~J.}\ \bibnamefont
  {Schoelkopf}}\ and\ \bibinfo {author} {\bibfnamefont {S.~M.}\ \bibnamefont
  {Girvin}},\ }\href@noop {} {\bibfield  {journal} {\bibinfo  {journal}
  {Nature}\ }\textbf {\bibinfo {volume} {451}},\ \bibinfo {pages} {664}
  (\bibinfo {year} {2008})}\BibitemShut {NoStop}%
\bibitem [{\citenamefont {Niemczyk}\ \emph {et~al.}(2010)\citenamefont
  {Niemczyk}, \citenamefont {Deppe}, \citenamefont {Huebl}, \citenamefont
  {Menzel}, \citenamefont {Hocke}, \citenamefont {Schwarz}, \citenamefont
  {Garcia-Ripoll}, \citenamefont {Zueco}, \citenamefont {H{\"u}mmer},
  \citenamefont {Solano}, \citenamefont {Marx},\ and\ \citenamefont
  {Gross}}]{Niemczyk:2010gv}%
  \BibitemOpen
  \bibfield  {author} {\bibinfo {author} {\bibfnamefont {T.}~\bibnamefont
  {Niemczyk}}, \bibinfo {author} {\bibfnamefont {F.}~\bibnamefont {Deppe}},
  \bibinfo {author} {\bibfnamefont {H.}~\bibnamefont {Huebl}}, \bibinfo
  {author} {\bibfnamefont {E.~P.}\ \bibnamefont {Menzel}}, \bibinfo {author}
  {\bibfnamefont {F.}~\bibnamefont {Hocke}}, \bibinfo {author} {\bibfnamefont
  {M.~J.}\ \bibnamefont {Schwarz}}, \bibinfo {author} {\bibfnamefont {J.~J.}\
  \bibnamefont {Garcia-Ripoll}}, \bibinfo {author} {\bibfnamefont
  {D.}~\bibnamefont {Zueco}}, \bibinfo {author} {\bibfnamefont
  {T.}~\bibnamefont {H{\"u}mmer}}, \bibinfo {author} {\bibfnamefont
  {E.}~\bibnamefont {Solano}}, \bibinfo {author} {\bibfnamefont
  {A.}~\bibnamefont {Marx}}, \ and\ \bibinfo {author} {\bibfnamefont
  {R.}~\bibnamefont {Gross}},\ }\href@noop {} {\bibfield  {journal} {\bibinfo
  {journal} {Nat. Phys.}\ }\textbf {\bibinfo {volume} {6}},\ \bibinfo {pages}
  {772} (\bibinfo {year} {2010})}\BibitemShut {NoStop}%
\bibitem [{\citenamefont {Andr\'e}\ \emph {et~al.}(2006)\citenamefont
  {Andr\'e}, \citenamefont {DeMille}, \citenamefont {Doyle}, \citenamefont
  {Lukin}, \citenamefont {Maxwell}, \citenamefont {Rabl}, \citenamefont
  {Schoelkopf},\ and\ \citenamefont {Zoller}}]{Andre:2006}%
  \BibitemOpen
  \bibfield  {author} {\bibinfo {author} {\bibfnamefont {A.}~\bibnamefont
  {Andr\'e}}, \bibinfo {author} {\bibfnamefont {D.}~\bibnamefont {DeMille}},
  \bibinfo {author} {\bibfnamefont {J.~M.}\ \bibnamefont {Doyle}}, \bibinfo
  {author} {\bibfnamefont {M.~D.}\ \bibnamefont {Lukin}}, \bibinfo {author}
  {\bibfnamefont {S.~E.}\ \bibnamefont {Maxwell}}, \bibinfo {author}
  {\bibfnamefont {P.}~\bibnamefont {Rabl}}, \bibinfo {author} {\bibfnamefont
  {R.~J.}\ \bibnamefont {Schoelkopf}}, \ and\ \bibinfo {author} {\bibfnamefont
  {P.}~\bibnamefont {Zoller}},\ }\href@noop {} {\bibfield  {journal} {\bibinfo
  {journal} {Nat. Phys.}\ }\textbf {\bibinfo {volume} {2}},\ \bibinfo {pages}
  {636} (\bibinfo {year} {2006})}\BibitemShut {NoStop}%
\bibitem [{\citenamefont {Verd\'u}\ \emph {et~al.}(2009)\citenamefont
  {Verd\'u}, \citenamefont {Zoubi}, \citenamefont {Koller}, \citenamefont
  {Majer}, \citenamefont {Ritsch},\ and\ \citenamefont
  {Schmiedmayer}}]{Verdu:2009}%
  \BibitemOpen
  \bibfield  {author} {\bibinfo {author} {\bibfnamefont {J.}~\bibnamefont
  {Verd\'u}}, \bibinfo {author} {\bibfnamefont {H.}~\bibnamefont {Zoubi}},
  \bibinfo {author} {\bibfnamefont {C.}~\bibnamefont {Koller}}, \bibinfo
  {author} {\bibfnamefont {J.}~\bibnamefont {Majer}}, \bibinfo {author}
  {\bibfnamefont {H.}~\bibnamefont {Ritsch}}, \ and\ \bibinfo {author}
  {\bibfnamefont {J.}~\bibnamefont {Schmiedmayer}},\ }\href {\doibase
  10.1103/PhysRevLett.103.043603} {\bibfield  {journal} {\bibinfo  {journal}
  {Phys. Rev. Lett.}\ }\textbf {\bibinfo {volume} {103}},\ \bibinfo {pages}
  {043603} (\bibinfo {year} {2009})}\BibitemShut {NoStop}%
\bibitem [{\citenamefont {Wallquist}\ \emph {et~al.}(2009)\citenamefont
  {Wallquist}, \citenamefont {Hammerer}, \citenamefont {Rabl}, \citenamefont
  {Lukin},\ and\ \citenamefont {Zoller}}]{Wallquist:2009}%
  \BibitemOpen
  \bibfield  {author} {\bibinfo {author} {\bibfnamefont {M.}~\bibnamefont
  {Wallquist}}, \bibinfo {author} {\bibfnamefont {K.}~\bibnamefont {Hammerer}},
  \bibinfo {author} {\bibfnamefont {P.}~\bibnamefont {Rabl}}, \bibinfo {author}
  {\bibfnamefont {M.}~\bibnamefont {Lukin}}, \ and\ \bibinfo {author}
  {\bibfnamefont {P.}~\bibnamefont {Zoller}},\ }\href@noop {} {\bibfield
  {journal} {\bibinfo  {journal} {Phys. Scr.}\ }\textbf {\bibinfo {volume}
  {T137}},\ \bibinfo {pages} {014001} (\bibinfo {year} {2009})}\BibitemShut
  {NoStop}%
\bibitem [{\citenamefont {DiCarlo}\ \emph {et~al.}(2009)\citenamefont
  {DiCarlo}, \citenamefont {Chow}, \citenamefont {Gambetta}, \citenamefont
  {Bishop}, \citenamefont {Johnson}, \citenamefont {Schuster}, \citenamefont
  {Majer}, \citenamefont {Blais}, \citenamefont {Frunzio}, \citenamefont
  {Girvin},\ and\ \citenamefont {Schoelkopf}}]{dicarlo09}%
  \BibitemOpen
  \bibfield  {author} {\bibinfo {author} {\bibfnamefont {L.}~\bibnamefont
  {DiCarlo}}, \bibinfo {author} {\bibfnamefont {J.~M.}\ \bibnamefont {Chow}},
  \bibinfo {author} {\bibfnamefont {J.~M.}\ \bibnamefont {Gambetta}}, \bibinfo
  {author} {\bibfnamefont {L.~S.}\ \bibnamefont {Bishop}}, \bibinfo {author}
  {\bibfnamefont {B.~R.}\ \bibnamefont {Johnson}}, \bibinfo {author}
  {\bibfnamefont {D.~I.}\ \bibnamefont {Schuster}}, \bibinfo {author}
  {\bibfnamefont {J.}~\bibnamefont {Majer}}, \bibinfo {author} {\bibfnamefont
  {A.}~\bibnamefont {Blais}}, \bibinfo {author} {\bibfnamefont
  {L.}~\bibnamefont {Frunzio}}, \bibinfo {author} {\bibfnamefont {S.~M.}\
  \bibnamefont {Girvin}}, \ and\ \bibinfo {author} {\bibfnamefont {R.~J.}\
  \bibnamefont {Schoelkopf}},\ }\href@noop {} {\bibfield  {journal} {\bibinfo
  {journal} {Nature}\ }\textbf {\bibinfo {volume} {460}},\ \bibinfo {pages}
  {240} (\bibinfo {year} {2009})}\BibitemShut {NoStop}%
\bibitem [{\citenamefont {Wesenberg}\ \emph {et~al.}(2009)\citenamefont
  {Wesenberg}, \citenamefont {Ardavan}, \citenamefont {Briggs}, \citenamefont
  {Morton}, \citenamefont {Schoelkopf}, \citenamefont {Schuster},\ and\
  \citenamefont {Molmer}}]{Wesenberg:2009es}%
  \BibitemOpen
  \bibfield  {author} {\bibinfo {author} {\bibfnamefont {J.~H.}\ \bibnamefont
  {Wesenberg}}, \bibinfo {author} {\bibfnamefont {A.}~\bibnamefont {Ardavan}},
  \bibinfo {author} {\bibfnamefont {G.~A.~D.}\ \bibnamefont {Briggs}}, \bibinfo
  {author} {\bibfnamefont {J.~J.~L.}\ \bibnamefont {Morton}}, \bibinfo {author}
  {\bibfnamefont {R.~J.}\ \bibnamefont {Schoelkopf}}, \bibinfo {author}
  {\bibfnamefont {D.~I.}\ \bibnamefont {Schuster}}, \ and\ \bibinfo {author}
  {\bibfnamefont {K.}~\bibnamefont {Molmer}},\ }\href@noop {} {\bibfield
  {journal} {\bibinfo  {journal} {Phys. Rev. Lett.}\ }\textbf {\bibinfo
  {volume} {103}},\ \bibinfo {pages} {070502} (\bibinfo {year}
  {2009})}\BibitemShut {NoStop}%
\bibitem [{\citenamefont {Imamoglu}(2009)}]{Imamoglu:2009eg}%
  \BibitemOpen
  \bibfield  {author} {\bibinfo {author} {\bibfnamefont {A.}~\bibnamefont
  {Imamoglu}},\ }\href@noop {} {\bibfield  {journal} {\bibinfo  {journal}
  {Phys. Rev. Lett.}\ }\textbf {\bibinfo {volume} {102}},\ \bibinfo {pages}
  {083602} (\bibinfo {year} {2009})}\BibitemShut {NoStop}%
\bibitem [{\citenamefont {Schuster}\ \emph {et~al.}(2010)\citenamefont
  {Schuster}, \citenamefont {Sears}, \citenamefont {Ginossar}, \citenamefont
  {DiCarlo}, \citenamefont {Frunzio}, \citenamefont {Morton}, \citenamefont
  {Wu}, \citenamefont {Briggs}, \citenamefont {Buckley}, \citenamefont
  {Awschalom},\ and\ \citenamefont {Schoelkopf}}]{Schuster:2010bu}%
  \BibitemOpen
  \bibfield  {author} {\bibinfo {author} {\bibfnamefont {D.~I.}\ \bibnamefont
  {Schuster}}, \bibinfo {author} {\bibfnamefont {A.~P.}\ \bibnamefont {Sears}},
  \bibinfo {author} {\bibfnamefont {E.}~\bibnamefont {Ginossar}}, \bibinfo
  {author} {\bibfnamefont {L.}~\bibnamefont {DiCarlo}}, \bibinfo {author}
  {\bibfnamefont {L.}~\bibnamefont {Frunzio}}, \bibinfo {author} {\bibfnamefont
  {J.~J.~L.}\ \bibnamefont {Morton}}, \bibinfo {author} {\bibfnamefont
  {H.}~\bibnamefont {Wu}}, \bibinfo {author} {\bibfnamefont {G.~A.~D.}\
  \bibnamefont {Briggs}}, \bibinfo {author} {\bibfnamefont {B.~B.}\
  \bibnamefont {Buckley}}, \bibinfo {author} {\bibfnamefont {D.~D.}\
  \bibnamefont {Awschalom}}, \ and\ \bibinfo {author} {\bibfnamefont {R.~J.}\
  \bibnamefont {Schoelkopf}},\ }\href@noop {} {\bibfield  {journal} {\bibinfo
  {journal} {Phys. Rev. Lett.}\ }\textbf {\bibinfo {volume} {105}},\ \bibinfo
  {pages} {140501} (\bibinfo {year} {2010})}\BibitemShut {NoStop}%
\bibitem [{\citenamefont {Kubo}\ \emph {et~al.}(2010)\citenamefont {Kubo},
  \citenamefont {Ong}, \citenamefont {Bertet}, \citenamefont {Vion},
  \citenamefont {Jacques}, \citenamefont {Zheng}, \citenamefont {Dreau},
  \citenamefont {Roch}, \citenamefont {Auffeves}, \citenamefont {Jelezko},
  \citenamefont {Wrachtrup}, \citenamefont {Barthe}, \citenamefont {Bergonzo},\
  and\ \citenamefont {Esteve}}]{Kubo:2010iq}%
  \BibitemOpen
  \bibfield  {author} {\bibinfo {author} {\bibfnamefont {Y.}~\bibnamefont
  {Kubo}}, \bibinfo {author} {\bibfnamefont {F.~R.}\ \bibnamefont {Ong}},
  \bibinfo {author} {\bibfnamefont {P.}~\bibnamefont {Bertet}}, \bibinfo
  {author} {\bibfnamefont {D.}~\bibnamefont {Vion}}, \bibinfo {author}
  {\bibfnamefont {V.}~\bibnamefont {Jacques}}, \bibinfo {author} {\bibfnamefont
  {D.}~\bibnamefont {Zheng}}, \bibinfo {author} {\bibfnamefont
  {A.}~\bibnamefont {Dreau}}, \bibinfo {author} {\bibfnamefont {J.~F.}\
  \bibnamefont {Roch}}, \bibinfo {author} {\bibfnamefont {A.}~\bibnamefont
  {Auffeves}}, \bibinfo {author} {\bibfnamefont {F.}~\bibnamefont {Jelezko}},
  \bibinfo {author} {\bibfnamefont {J.}~\bibnamefont {Wrachtrup}}, \bibinfo
  {author} {\bibfnamefont {M.~F.}\ \bibnamefont {Barthe}}, \bibinfo {author}
  {\bibfnamefont {P.}~\bibnamefont {Bergonzo}}, \ and\ \bibinfo {author}
  {\bibfnamefont {D.}~\bibnamefont {Esteve}},\ }\href@noop {} {\bibfield
  {journal} {\bibinfo  {journal} {Phys. Rev. Lett.}\ }\textbf {\bibinfo
  {volume} {105}},\ \bibinfo {pages} {140502} (\bibinfo {year}
  {2010})}\BibitemShut {NoStop}%
\bibitem [{\citenamefont {Kubo}\ \emph {et~al.}(2011)\citenamefont {Kubo},
  \citenamefont {Grezes}, \citenamefont {Dewes}, \citenamefont {Umeda},
  \citenamefont {Isoya}, \citenamefont {Sumiya}, \citenamefont {Morishita},
  \citenamefont {Abe}, \citenamefont {Onoda}, \citenamefont {Ohshima},
  \citenamefont {Jacques}, \citenamefont {Dr\'eau}, \citenamefont {Roch},
  \citenamefont {Diniz}, \citenamefont {Auffeves}, \citenamefont {Vion},
  \citenamefont {Esteve},\ and\ \citenamefont {Bertet}}]{Kubo:2011dx}%
  \BibitemOpen
  \bibfield  {author} {\bibinfo {author} {\bibfnamefont {Y.}~\bibnamefont
  {Kubo}}, \bibinfo {author} {\bibfnamefont {C.}~\bibnamefont {Grezes}},
  \bibinfo {author} {\bibfnamefont {A.}~\bibnamefont {Dewes}}, \bibinfo
  {author} {\bibfnamefont {T.}~\bibnamefont {Umeda}}, \bibinfo {author}
  {\bibfnamefont {J.}~\bibnamefont {Isoya}}, \bibinfo {author} {\bibfnamefont
  {H.}~\bibnamefont {Sumiya}}, \bibinfo {author} {\bibfnamefont
  {N.}~\bibnamefont {Morishita}}, \bibinfo {author} {\bibfnamefont
  {H.}~\bibnamefont {Abe}}, \bibinfo {author} {\bibfnamefont {S.}~\bibnamefont
  {Onoda}}, \bibinfo {author} {\bibfnamefont {T.}~\bibnamefont {Ohshima}},
  \bibinfo {author} {\bibfnamefont {V.}~\bibnamefont {Jacques}}, \bibinfo
  {author} {\bibfnamefont {A.}~\bibnamefont {Dr\'eau}}, \bibinfo {author}
  {\bibfnamefont {J.-F.}\ \bibnamefont {Roch}}, \bibinfo {author}
  {\bibfnamefont {I.}~\bibnamefont {Diniz}}, \bibinfo {author} {\bibfnamefont
  {A.}~\bibnamefont {Auffeves}}, \bibinfo {author} {\bibfnamefont
  {D.}~\bibnamefont {Vion}}, \bibinfo {author} {\bibfnamefont {D.}~\bibnamefont
  {Esteve}}, \ and\ \bibinfo {author} {\bibfnamefont {P.}~\bibnamefont
  {Bertet}},\ }\href {\doibase 10.1103/PhysRevLett.107.220501} {\bibfield
  {journal} {\bibinfo  {journal} {Phys. Rev. Lett.}\ }\textbf {\bibinfo
  {volume} {107}},\ \bibinfo {pages} {220501} (\bibinfo {year}
  {2011})}\BibitemShut {NoStop}%
\bibitem [{\citenamefont {Ams\"uss}\ \emph {et~al.}(2011)\citenamefont
  {Ams\"uss}, \citenamefont {Koller}, \citenamefont {N\"obauer}, \citenamefont
  {Putz}, \citenamefont {Rotter}, \citenamefont {Sandner}, \citenamefont
  {Schneider}, \citenamefont {Schramb\"ock}, \citenamefont {Steinhauser},
  \citenamefont {Ritsch}, \citenamefont {Schmiedmayer},\ and\ \citenamefont
  {Majer}}]{Amsuss:2011ci}%
  \BibitemOpen
  \bibfield  {author} {\bibinfo {author} {\bibfnamefont {R.}~\bibnamefont
  {Ams\"uss}}, \bibinfo {author} {\bibfnamefont {C.}~\bibnamefont {Koller}},
  \bibinfo {author} {\bibfnamefont {T.}~\bibnamefont {N\"obauer}}, \bibinfo
  {author} {\bibfnamefont {S.}~\bibnamefont {Putz}}, \bibinfo {author}
  {\bibfnamefont {S.}~\bibnamefont {Rotter}}, \bibinfo {author} {\bibfnamefont
  {K.}~\bibnamefont {Sandner}}, \bibinfo {author} {\bibfnamefont
  {S.}~\bibnamefont {Schneider}}, \bibinfo {author} {\bibfnamefont
  {M.}~\bibnamefont {Schramb\"ock}}, \bibinfo {author} {\bibfnamefont
  {G.}~\bibnamefont {Steinhauser}}, \bibinfo {author} {\bibfnamefont
  {H.}~\bibnamefont {Ritsch}}, \bibinfo {author} {\bibfnamefont
  {J.}~\bibnamefont {Schmiedmayer}}, \ and\ \bibinfo {author} {\bibfnamefont
  {J.}~\bibnamefont {Majer}},\ }\href {\doibase 10.1103/PhysRevLett.107.060502}
  {\bibfield  {journal} {\bibinfo  {journal} {Phys. Rev. Lett.}\ }\textbf
  {\bibinfo {volume} {107}},\ \bibinfo {pages} {060502} (\bibinfo {year}
  {2011})}\BibitemShut {NoStop}%
\bibitem [{\citenamefont {Chiorescu}\ \emph {et~al.}(2010)\citenamefont
  {Chiorescu}, \citenamefont {Groll}, \citenamefont {Bertaina}, \citenamefont
  {Mori},\ and\ \citenamefont {Miyashita}}]{Chiorescu:2010hw}%
  \BibitemOpen
  \bibfield  {author} {\bibinfo {author} {\bibfnamefont {I.}~\bibnamefont
  {Chiorescu}}, \bibinfo {author} {\bibfnamefont {N.}~\bibnamefont {Groll}},
  \bibinfo {author} {\bibfnamefont {S.}~\bibnamefont {Bertaina}}, \bibinfo
  {author} {\bibfnamefont {T.}~\bibnamefont {Mori}}, \ and\ \bibinfo {author}
  {\bibfnamefont {S.}~\bibnamefont {Miyashita}},\ }\href@noop {} {\bibfield
  {journal} {\bibinfo  {journal} {Phys. Rev. B}\ }\textbf {\bibinfo {volume}
  {82}},\ \bibinfo {pages} {024413} (\bibinfo {year} {2010})}\BibitemShut
  {NoStop}%
\bibitem [{\citenamefont {Bushev}\ \emph {et~al.}(2011)\citenamefont {Bushev},
  \citenamefont {Feofanov}, \citenamefont {Rotzinger}, \citenamefont
  {Protopopov}, \citenamefont {Cole}, \citenamefont {Wilson}, \citenamefont
  {Fischer}, \citenamefont {Lukashenko},\ and\ \citenamefont
  {Ustinov}}]{Bushev:2011be}%
  \BibitemOpen
  \bibfield  {author} {\bibinfo {author} {\bibfnamefont {P.}~\bibnamefont
  {Bushev}}, \bibinfo {author} {\bibfnamefont {A.~K.}\ \bibnamefont
  {Feofanov}}, \bibinfo {author} {\bibfnamefont {H.}~\bibnamefont {Rotzinger}},
  \bibinfo {author} {\bibfnamefont {I.}~\bibnamefont {Protopopov}}, \bibinfo
  {author} {\bibfnamefont {J.~H.}\ \bibnamefont {Cole}}, \bibinfo {author}
  {\bibfnamefont {C.~M.}\ \bibnamefont {Wilson}}, \bibinfo {author}
  {\bibfnamefont {G.}~\bibnamefont {Fischer}}, \bibinfo {author} {\bibfnamefont
  {A.}~\bibnamefont {Lukashenko}}, \ and\ \bibinfo {author} {\bibfnamefont
  {A.~V.}\ \bibnamefont {Ustinov}},\ }\href {\doibase
  10.1103/PhysRevB.84.060501} {\bibfield  {journal} {\bibinfo  {journal} {Phys.
  Rev. B}\ }\textbf {\bibinfo {volume} {84}},\ \bibinfo {pages} {060501}
  (\bibinfo {year} {2011})}\BibitemShut {NoStop}%
\bibitem [{\citenamefont {Abe}\ \emph {et~al.}(2011)\citenamefont {Abe},
  \citenamefont {Wu}, \citenamefont {Ardavan},\ and\ \citenamefont
  {Morton}}]{Abe:2011gv}%
  \BibitemOpen
  \bibfield  {author} {\bibinfo {author} {\bibfnamefont {E.}~\bibnamefont
  {Abe}}, \bibinfo {author} {\bibfnamefont {H.}~\bibnamefont {Wu}}, \bibinfo
  {author} {\bibfnamefont {A.}~\bibnamefont {Ardavan}}, \ and\ \bibinfo
  {author} {\bibfnamefont {J.~J.~L.}\ \bibnamefont {Morton}},\ }\href@noop {}
  {\bibfield  {journal} {\bibinfo  {journal} {Appl. Phys. Lett.}\ }\textbf
  {\bibinfo {volume} {98}},\ \bibinfo {pages} {251108} (\bibinfo {year}
  {2011})}\BibitemShut {NoStop}%
\bibitem [{\citenamefont {Zhu}\ \emph {et~al.}(2011)\citenamefont {Zhu},
  \citenamefont {Saito}, \citenamefont {Kemp}, \citenamefont {Kakuyanagi},
  \citenamefont {Karimoto}, \citenamefont {Nakano}, \citenamefont {Munro},
  \citenamefont {Tokura}, \citenamefont {Everitt}, \citenamefont {Nemoto},
  \citenamefont {Kasu}, \citenamefont {Mizuochi},\ and\ \citenamefont
  {Semba}}]{Zhu:2011}%
  \BibitemOpen
  \bibfield  {author} {\bibinfo {author} {\bibfnamefont {X.}~\bibnamefont
  {Zhu}}, \bibinfo {author} {\bibfnamefont {S.}~\bibnamefont {Saito}}, \bibinfo
  {author} {\bibfnamefont {A.}~\bibnamefont {Kemp}}, \bibinfo {author}
  {\bibfnamefont {K.}~\bibnamefont {Kakuyanagi}}, \bibinfo {author}
  {\bibfnamefont {S.}~\bibnamefont {Karimoto}}, \bibinfo {author}
  {\bibfnamefont {H.}~\bibnamefont {Nakano}}, \bibinfo {author} {\bibfnamefont
  {W.~J.}\ \bibnamefont {Munro}}, \bibinfo {author} {\bibfnamefont
  {Y.}~\bibnamefont {Tokura}}, \bibinfo {author} {\bibfnamefont {M.~S.}\
  \bibnamefont {Everitt}}, \bibinfo {author} {\bibfnamefont {K.}~\bibnamefont
  {Nemoto}}, \bibinfo {author} {\bibfnamefont {M.}~\bibnamefont {Kasu}},
  \bibinfo {author} {\bibfnamefont {N.}~\bibnamefont {Mizuochi}}, \ and\
  \bibinfo {author} {\bibfnamefont {K.}~\bibnamefont {Semba}},\ }\href@noop {}
  {\bibfield  {journal} {\bibinfo  {journal} {Nature}\ }\textbf {\bibinfo
  {volume} {478}},\ \bibinfo {pages} {221} (\bibinfo {year}
  {2011})}\BibitemShut {NoStop}%
\bibitem [{\citenamefont {Raizen}\ \emph {et~al.}(1989)\citenamefont {Raizen},
  \citenamefont {Thompson}, \citenamefont {Brecha}, \citenamefont {Kimble},\
  and\ \citenamefont {Carmichael}}]{Raizen:1989}%
  \BibitemOpen
  \bibfield  {author} {\bibinfo {author} {\bibfnamefont {M.~G.}\ \bibnamefont
  {Raizen}}, \bibinfo {author} {\bibfnamefont {R.~J.}\ \bibnamefont
  {Thompson}}, \bibinfo {author} {\bibfnamefont {R.~J.}\ \bibnamefont
  {Brecha}}, \bibinfo {author} {\bibfnamefont {H.~J.}\ \bibnamefont {Kimble}},
  \ and\ \bibinfo {author} {\bibfnamefont {H.~J.}\ \bibnamefont {Carmichael}},\
  }\href {\doibase 10.1103/PhysRevLett.63.240} {\bibfield  {journal} {\bibinfo
  {journal} {Phys. Rev. Lett.}\ }\textbf {\bibinfo {volume} {63}},\ \bibinfo
  {pages} {240} (\bibinfo {year} {1989})}\BibitemShut {NoStop}%
\bibitem [{\citenamefont {Walls}\ and\ \citenamefont
  {Milburn}(1994)}]{wallsmilburn94}%
  \BibitemOpen
  \bibfield  {author} {\bibinfo {author} {\bibfnamefont {D.~F.}\ \bibnamefont
  {Walls}}\ and\ \bibinfo {author} {\bibfnamefont {G.~J.}\ \bibnamefont
  {Milburn}},\ }\href@noop {} {\emph {\bibinfo {title} {{Quantum Optics}}}},\
  \bibinfo {edition} {1st}\ ed.\ (\bibinfo  {publisher} {Springer},\ \bibinfo
  {address} {Berlin},\ \bibinfo {year} {1994})\BibitemShut {NoStop}%
\bibitem [{\citenamefont {Soykal}\ and\ \citenamefont
  {Flatt\'e}(2010{\natexlab{a}})}]{Soykal:2010er}%
  \BibitemOpen
  \bibfield  {author} {\bibinfo {author} {\bibfnamefont {O.~O.}\ \bibnamefont
  {Soykal}}\ and\ \bibinfo {author} {\bibfnamefont {M.~E.}\ \bibnamefont
  {Flatt\'e}},\ }\href {\doibase 10.1103/PhysRevB.82.104413} {\bibfield
  {journal} {\bibinfo  {journal} {Phys. Rev. B}\ }\textbf {\bibinfo {volume}
  {82}},\ \bibinfo {pages} {104413} (\bibinfo {year}
  {2010}{\natexlab{a}})}\BibitemShut {NoStop}%
\bibitem [{\citenamefont {Soykal}\ and\ \citenamefont
  {Flatt\'e}(2010{\natexlab{b}})}]{Soykal:2010hz}%
  \BibitemOpen
  \bibfield  {author} {\bibinfo {author} {\bibfnamefont {O.~O.}\ \bibnamefont
  {Soykal}}\ and\ \bibinfo {author} {\bibfnamefont {M.~E.}\ \bibnamefont
  {Flatt\'e}},\ }\href {\doibase 10.1103/PhysRevLett.104.077202} {\bibfield
  {journal} {\bibinfo  {journal} {Phys. Rev. Lett.}\ }\textbf {\bibinfo
  {volume} {104}},\ \bibinfo {pages} {077202} (\bibinfo {year}
  {2010}{\natexlab{b}})}\BibitemShut {NoStop}%
\bibitem [{\citenamefont {Rachford}\ \emph {et~al.}(2000)\citenamefont
  {Rachford}, \citenamefont {Levy}, \citenamefont {Osgood}, \citenamefont
  {Kumar},\ and\ \citenamefont {Bakhru}}]{Rachford:2000en}%
  \BibitemOpen
  \bibfield  {author} {\bibinfo {author} {\bibfnamefont {F.~J.}\ \bibnamefont
  {Rachford}}, \bibinfo {author} {\bibfnamefont {M.}~\bibnamefont {Levy}},
  \bibinfo {author} {\bibfnamefont {R.~M.}\ \bibnamefont {Osgood}}, \bibinfo
  {author} {\bibfnamefont {A.}~\bibnamefont {Kumar}}, \ and\ \bibinfo {author}
  {\bibfnamefont {H.}~\bibnamefont {Bakhru}},\ }\href@noop {} {\bibfield
  {journal} {\bibinfo  {journal} {J. Appl. Phys.}\ }\textbf {\bibinfo {volume}
  {87}},\ \bibinfo {pages} {6253} (\bibinfo {year} {2000})}\BibitemShut
  {NoStop}%
\bibitem [{\citenamefont {Spencer}\ \emph {et~al.}(1959)\citenamefont
  {Spencer}, \citenamefont {LeCraw},\ and\ \citenamefont
  {Clogston}}]{SPENCER:1959wv}%
  \BibitemOpen
  \bibfield  {author} {\bibinfo {author} {\bibfnamefont {E.~G.}\ \bibnamefont
  {Spencer}}, \bibinfo {author} {\bibfnamefont {R.~C.}\ \bibnamefont {LeCraw}},
  \ and\ \bibinfo {author} {\bibfnamefont {A.~M.}\ \bibnamefont {Clogston}},\
  }\href@noop {} {\bibfield  {journal} {\bibinfo  {journal} {Phys. Rev. Lett.}\
  }\textbf {\bibinfo {volume} {3}},\ \bibinfo {pages} {32} (\bibinfo {year}
  {1959})}\BibitemShut {NoStop}%
\bibitem [{\citenamefont {Kaplan}(1965)}]{Kaplan:1965hf}%
  \BibitemOpen
  \bibfield  {author} {\bibinfo {author} {\bibfnamefont {D.~E.}\ \bibnamefont
  {Kaplan}},\ }\href@noop {} {\bibfield  {journal} {\bibinfo  {journal} {Phys.
  Rev. Lett.}\ }\textbf {\bibinfo {volume} {14}},\ \bibinfo {pages} {254}
  (\bibinfo {year} {1965})}\BibitemShut {NoStop}%
\bibitem [{\citenamefont {Manuilov}\ \emph {et~al.}(2009)\citenamefont
  {Manuilov}, \citenamefont {Fors}, \citenamefont {Khartsev},\ and\
  \citenamefont {Grishin}}]{Manuilov:2009kl}%
  \BibitemOpen
  \bibfield  {author} {\bibinfo {author} {\bibfnamefont {S.~A.}\ \bibnamefont
  {Manuilov}}, \bibinfo {author} {\bibfnamefont {R.}~\bibnamefont {Fors}},
  \bibinfo {author} {\bibfnamefont {S.~I.}\ \bibnamefont {Khartsev}}, \ and\
  \bibinfo {author} {\bibfnamefont {A.~M.}\ \bibnamefont {Grishin}},\
  }\href@noop {} {\bibfield  {journal} {\bibinfo  {journal} {J. Appl. Phys.}\
  }\textbf {\bibinfo {volume} {105}},\ \bibinfo {pages} {033917} (\bibinfo
  {year} {2009})}\BibitemShut {NoStop}%
\bibitem [{\citenamefont {Manuilov}\ and\ \citenamefont
  {Grishin}(2010)}]{Manuilov:2010ix}%
  \BibitemOpen
  \bibfield  {author} {\bibinfo {author} {\bibfnamefont {S.~A.}\ \bibnamefont
  {Manuilov}}\ and\ \bibinfo {author} {\bibfnamefont {A.~M.}\ \bibnamefont
  {Grishin}},\ }\href@noop {} {\bibfield  {journal} {\bibinfo  {journal} {J.
  Appl. Phys.}\ }\textbf {\bibinfo {volume} {108}},\ \bibinfo {pages} {013902}
  (\bibinfo {year} {2010})}\BibitemShut {NoStop}%
\bibitem [{\citenamefont {Levy}\ \emph {et~al.}(1997)\citenamefont {Levy},
  \citenamefont {Osgood}, \citenamefont {Kumar},\ and\ \citenamefont
  {Bakhru}}]{Levy:1997ki}%
  \BibitemOpen
  \bibfield  {author} {\bibinfo {author} {\bibfnamefont {M.}~\bibnamefont
  {Levy}}, \bibinfo {author} {\bibfnamefont {R.~M.}\ \bibnamefont {Osgood}},
  \bibinfo {author} {\bibfnamefont {A.}~\bibnamefont {Kumar}}, \ and\ \bibinfo
  {author} {\bibfnamefont {H.}~\bibnamefont {Bakhru}},\ }\href@noop {}
  {\bibfield  {journal} {\bibinfo  {journal} {Appl. Phys. Lett.}\ }\textbf
  {\bibinfo {volume} {71}},\ \bibinfo {pages} {2617} (\bibinfo {year}
  {1997})}\BibitemShut {NoStop}%
\bibitem [{Note1()}]{Note1}%
  \BibitemOpen
  \bibinfo {note} {The magnetic induction $B_z^{\protect \rm
  {eff}}=B_z^{\protect \rm {ext}}+B_{\protect \rm {a}}$ has two contributions:
  i) the externally applied magnetic field $B_z^{\protect \rm {ext}}$ and ii)
  internal fields accounting for the magnetic anisotropy $B_{\protect \rm
  {a}}$. (cf. \cite {SI})}\BibitemShut {NoStop}%
\bibitem [{\citenamefont {Dicke}(1954)}]{Dicke:1954}%
  \BibitemOpen
  \bibfield  {author} {\bibinfo {author} {\bibfnamefont {R.~H.}\ \bibnamefont
  {Dicke}},\ }\href {\doibase 10.1103/PhysRev.93.99} {\bibfield  {journal}
  {\bibinfo  {journal} {Phys. Rev.}\ }\textbf {\bibinfo {volume} {93}},\
  \bibinfo {pages} {99} (\bibinfo {year} {1954})}\BibitemShut {NoStop}%
\bibitem [{\citenamefont {Niemczyk}\ \emph {et~al.}(2009)\citenamefont
  {Niemczyk}, \citenamefont {Deppe}, \citenamefont {Mariantoni}, \citenamefont
  {Menzel}, \citenamefont {Hoffmann}, \citenamefont {Wild}, \citenamefont
  {Eggenstein}, \citenamefont {Marx},\ and\ \citenamefont
  {Gross}}]{niemczyk09}%
  \BibitemOpen
  \bibfield  {author} {\bibinfo {author} {\bibfnamefont {T.}~\bibnamefont
  {Niemczyk}}, \bibinfo {author} {\bibfnamefont {F.}~\bibnamefont {Deppe}},
  \bibinfo {author} {\bibfnamefont {M.}~\bibnamefont {Mariantoni}}, \bibinfo
  {author} {\bibfnamefont {E.~P.}\ \bibnamefont {Menzel}}, \bibinfo {author}
  {\bibfnamefont {E.}~\bibnamefont {Hoffmann}}, \bibinfo {author}
  {\bibfnamefont {G.}~\bibnamefont {Wild}}, \bibinfo {author} {\bibfnamefont
  {L.}~\bibnamefont {Eggenstein}}, \bibinfo {author} {\bibfnamefont
  {A.}~\bibnamefont {Marx}}, \ and\ \bibinfo {author} {\bibfnamefont
  {R.}~\bibnamefont {Gross}},\ }\href@noop {} {\bibfield  {journal} {\bibinfo
  {journal} {Supercond. Sci. Tech.}\ }\textbf {\bibinfo {volume} {22}},\
  \bibinfo {pages} {034009} (\bibinfo {year} {2009})}\BibitemShut {NoStop}%
\bibitem [{\citenamefont {Gilleo}\ and\ \citenamefont
  {Geller}(1958)}]{gilleo58}%
  \BibitemOpen
  \bibfield  {author} {\bibinfo {author} {\bibfnamefont {M.}~\bibnamefont
  {Gilleo}}\ and\ \bibinfo {author} {\bibfnamefont {S.}~\bibnamefont
  {Geller}},\ }\href@noop {} {\bibfield  {journal} {\bibinfo  {journal} {Phys.
  Rev.}\ }\textbf {\bibinfo {volume} {110}},\ \bibinfo {pages} {73} (\bibinfo
  {year} {1958})}\BibitemShut {NoStop}%
\bibitem [{Note2()}]{Note2}%
  \BibitemOpen
  \bibinfo {note} {This accounts for the fixed attenuators and does not include
  the lossy lines. For further details refer to \cite {SI}.}\BibitemShut
  {Stop}%
\bibitem [{Note3()}]{Note3}%
  \BibitemOpen
  \bibinfo {note} {Since we have not calibrated our setup at millikelvin
  temperatures, we show raw, uncalibrated, as-measured $S_{21}$ transmission
  data in Figs.\protect \,\ref {fig:spectra} and\protect \,\ref {fig:analysis}.
  In our opinion, this is the most honest way of presenting the data in lack of
  a proper full calibration. \label {footnote:uncalibrated}}\BibitemShut
  {NoStop}%
\bibitem [{SI()}]{SI}%
  \BibitemOpen
  \href@noop {} {\enquote {\bibinfo {title} {Sublemental material},}\
  }\BibitemShut {NoStop}%
\bibitem [{\citenamefont {Haroche}\ and\ \citenamefont
  {Raimond}(2006)}]{HarocheRaimond:2006}%
  \BibitemOpen
  \bibfield  {author} {\bibinfo {author} {\bibfnamefont {S.}~\bibnamefont
  {Haroche}}\ and\ \bibinfo {author} {\bibfnamefont {J.~M.}\ \bibnamefont
  {Raimond}},\ }\href@noop {} {\emph {\bibinfo {title} {{Exploring the Quantum:
  Atoms, Cavities and Photons}}}}\ (\bibinfo  {publisher} {Oxford University
  Press},\ \bibinfo {address} {Oxford},\ \bibinfo {year} {2006})\BibitemShut
  {NoStop}%
\bibitem [{\citenamefont {Belov}\ \emph {et~al.}(1960)\citenamefont {Belov},
  \citenamefont {Malevskaya},\ and\ \citenamefont {Sokolov}}]{Belov:1960va}%
  \BibitemOpen
  \bibfield  {author} {\bibinfo {author} {\bibfnamefont {K.~P.}\ \bibnamefont
  {Belov}}, \bibinfo {author} {\bibfnamefont {L.~A.}\ \bibnamefont
  {Malevskaya}}, \ and\ \bibinfo {author} {\bibfnamefont {V.~I.}\ \bibnamefont
  {Sokolov}},\ }\href@noop {} {\bibfield  {journal} {\bibinfo  {journal}
  {Soviet Physics JETP}\ }\textbf {\bibinfo {volume} {12}},\ \bibinfo {pages}
  {1074} (\bibinfo {year} {1960})}\BibitemShut {NoStop}%
\bibitem [{\citenamefont {Clerk}\ \emph {et~al.}(2010)\citenamefont {Clerk},
  \citenamefont {Girvin}, \citenamefont {Marquardt},\ and\ \citenamefont
  {Schoelkopf}}]{Clerk:2010dh}%
  \BibitemOpen
  \bibfield  {author} {\bibinfo {author} {\bibfnamefont {A.~A.}\ \bibnamefont
  {Clerk}}, \bibinfo {author} {\bibfnamefont {S.~M.}\ \bibnamefont {Girvin}},
  \bibinfo {author} {\bibfnamefont {F.}~\bibnamefont {Marquardt}}, \ and\
  \bibinfo {author} {\bibfnamefont {R.~J.}\ \bibnamefont {Schoelkopf}},\
  }\href@noop {} {\bibfield  {journal} {\bibinfo  {journal} {Rev. Mod. Phys.}\
  }\textbf {\bibinfo {volume} {82}},\ \bibinfo {pages} {1155} (\bibinfo {year}
  {2010})}\BibitemShut {NoStop}%
\bibitem [{Note4()}]{Note4}%
  \BibitemOpen
  \bibinfo {note} {Our claims are based on the analysis of the lineshape and
  resonance frequency dependence as function of the applied magnetic field and
  do not rely on the absolute transmission intensity. We therefore have not
  calibrated the setup with respect to the amplitude information.}\BibitemShut
  {Stop}%
\bibitem [{gay()}]{gayig-datasheet-2012}%
  \BibitemOpen
  \href {www.ferrisphere.com} {\emph {\bibinfo {title} {YIG:Ga Datasheet -
  www.ferrisphere.com}}}\BibitemShut {NoStop}%
\bibitem [{\citenamefont {Sparks}\ and\ \citenamefont
  {Kittel}(1960)}]{Sparks:1960}%
  \BibitemOpen
  \bibfield  {author} {\bibinfo {author} {\bibfnamefont {M.}~\bibnamefont
  {Sparks}}\ and\ \bibinfo {author} {\bibfnamefont {C.}~\bibnamefont
  {Kittel}},\ }\href {\doibase 10.1103/PhysRevLett.4.232} {\bibfield  {journal}
  {\bibinfo  {journal} {Phys. Rev. Lett.}\ }\textbf {\bibinfo {volume} {4}},\
  \bibinfo {pages} {232} (\bibinfo {year} {1960})}\BibitemShut {NoStop}%
\bibitem [{\citenamefont {Tyryshkin}\ \emph {et~al.}(2011)\citenamefont
  {Tyryshkin}, \citenamefont {Tojo}, \citenamefont {Morton}, \citenamefont
  {Riemann}, \citenamefont {Abrosimov}, \citenamefont {Becker}, \citenamefont
  {Pohl}, \citenamefont {Schenkel}, \citenamefont {Thewalt}, \citenamefont
  {Itoh},\ and\ \citenamefont {Lyon}}]{Tyryshkin:2011fi}%
  \BibitemOpen
  \bibfield  {author} {\bibinfo {author} {\bibfnamefont {A.~M.}\ \bibnamefont
  {Tyryshkin}}, \bibinfo {author} {\bibfnamefont {S.}~\bibnamefont {Tojo}},
  \bibinfo {author} {\bibfnamefont {J.~J.~L.}\ \bibnamefont {Morton}}, \bibinfo
  {author} {\bibfnamefont {H.}~\bibnamefont {Riemann}}, \bibinfo {author}
  {\bibfnamefont {N.~V.}\ \bibnamefont {Abrosimov}}, \bibinfo {author}
  {\bibfnamefont {P.}~\bibnamefont {Becker}}, \bibinfo {author} {\bibfnamefont
  {H.-J.}\ \bibnamefont {Pohl}}, \bibinfo {author} {\bibfnamefont
  {T.}~\bibnamefont {Schenkel}}, \bibinfo {author} {\bibfnamefont {M.~L.~W.}\
  \bibnamefont {Thewalt}}, \bibinfo {author} {\bibfnamefont {K.~M.}\
  \bibnamefont {Itoh}}, \ and\ \bibinfo {author} {\bibfnamefont {S.~A.}\
  \bibnamefont {Lyon}},\ }\href@noop {} {\bibfield  {journal} {\bibinfo
  {journal} {Nature Materials}\ }\textbf {\bibinfo {volume} {11}},\ \bibinfo
  {pages} {143} (\bibinfo {year} {2011})}\BibitemShut {NoStop}%
\end{thebibliography}
\end{document}